# Neutron diffraction study of a metallic kagome lattice, $Tb_3Ru_4Al_{12}$

**Sudhindra Rayaprol,**[1,*] **Andreas Hoser,**[2] **Kartik K Iyer,**[3] **Sanjay K Upadhyay,**[3]
**E. V. Sampathkumaran,**[3,*]

[1]*UGC-DAE-Consortium for Scientific Research, Mumbai Centre, BARC Campus, Trombay, Mumbai – 400085, India*
[2]*Helmholtz-Zentrum Berlin für Materialien und Energie, Berlin, D-14109, Germany*
[3]*Tata Institute of Fundamental Research, Homi Bhabha Road, Colaba, Mumbai 400005, India*

**Abstract**

We present the results of neutron diffraction studies on polycrystals of a metallic kagome lattice, $Tb_3Ru_4Al_{12}$, reported recently to undergo reentrant magnetism, with the onset of long range antiferromagnetic order below ($T_N$ = 22 K) and spin-glass features below about 17 K. The present results reveal long-range antiferromagnetic order of an incommensurate type with the moments oriented along *c*-axis at all temperatures below $T_N$. There are however notable changes in the *T*-dependence of propagation vector along *b*–axis across 17 K. An observation of interest is that there is no decrease of intensity of magnetic Bragg peaks on entrance into the glassy phase (that is, below 17 K). This finding suggests that the magnetism of this compound is an exotic one and one and we wonder whether this compound is an example for 'dynamical spin-glass freezing' phenomenon, as a consequence of geometrical frustration.

1. Introduction

In the field of magnetism, the materials containing magnetic ions forming triangular, tetrahedral and kagome lattices have been attracting a lot of attention due to geometrically frustrated magnetism in the event that the intersite interaction is antiferromagnetic to favour frustration. In this article, we focus on kagome lattices, which are generally characterized by magnetic ions placed at the corners of a hexagon with each face of the hexagon forming a triangle with another magnetic ion. Though the studies on such lattices are abundant among insulators with nearest neighbour antiferromagnetic interaction, such lattices are not commonly known in metallic environments, barring a few exceptions [1, 2]. In this respect, the rare-earth ($R$) compounds of the type, $R_3Ru_4Al_{12}$ [3-9], crystallizing in a hexagonal structure (space group, $P6_3/mmc$), containing $R$ ions in a distorted kagome network, provide an opportunity to study the phenomenon of geometrical frustration in the event that the magnetic ordering is mediated by Ruderman-Kittel-Kasuya-Yosida (RKKY) indirect exchange interaction. In this family, $Tb_3Ru_4Al_{12}$, known to order antiferromagnetically below ($T_N=$) 22 K [5], has been reported to show glassy characteristics at lower temperatures (<17 K) characterizing the compound as a re-entrant spin glass [9]. We have carried out neutron diffraction measurements at low temperatures in order to throw light on the magnetism of this compound.

2. **Experimental details**

The polycrystals for neutron diffraction measurements were freshly prepared as described in our previous studies [9] and characterized by x-ray diffraction. The sample was further characterized by the measurement of temperature ($T$) dependencies of dc magnetic susceptibility ($\chi$) and the results are in conformity with those reported earlier [9]. Neutron diffraction patterns on powdered sample were obtained (with the wavelength, $\lambda= 2.451$ Å) at several temperatures on E6 diffractometer at Helmholtz-Zentrum Berlin (HZB), Berlin. About 3.5 grams of powder sample was loaded in a vanadium can and attached to a sample holder insert, which was subsequently loaded into a standard liquid helium orange cryostat for measuring neutron diffraction patterns at several selected temperatures. The data for each spectrum was collected in 48 scan steps of the 2 two dimensional area detectors. The total angular range covered (in 2θ) is from 4 to 136 [10-12]. The specimen was first cooled to 4.5 K and the diffraction patterns were recorded at many temperatures while warming.

3. **Results and discussions**

In figure 1, we show the neutron diffraction patterns at selected temperatures below 28 K across $T_N$. The diffraction patterns were subjected to Rietveld fitting using FULLPROF suite programs [13, 14]. This fitting of the nuclear peaks (that is, above $T_N$) is consistent with $P6_3/mmc$ space group. The derived lattice constants are in good agreement with those reported in the literature [9] at room temperature, with a marginal decrease of unit-cell volume with decreasing temperature. The data confirm the existence of long (~5.19 Å) and short (~3.63 Å) Tb-Tb bond distances, and ~~hence~~ thus small and big triangles alternate within the kagome layer. As the temperature is lowered to 22 K, extra Bragg peaks (marked by vertical arrows in figure 1) tend to develop, which are attributable to the onset of long-range magnetic order. The fact that a strong peak appears at 2θ = ~ 6°, corresponding to a low Q = 0.26Å$^{-1}$ (Q= 4π sinθ / λ), suggests that the magnetic unit cell is larger than the crystallographic unit cell. Rietveld refined patterns plotted in figure 2 for $T$ = 4.5 K and 28 K show good agreement between observed and calculated model.

Details of the Rietveld analysis for nuclear and magnetic structures (along with magnetization data on the specimen employed for the present studies) are given in the Supplementary file [15]. The propagation vector, **k**, was obtained from the magnetic Bragg peak using the program, K-search (available in Winplot-R 2006 program, within Fullprof). Among various suggested **k** vectors, only the value, **k** = (0, ~1/3, 0) and the irreducible representations and basis vectors obtained using the program BasIreps for this **k**-vector, gave good fit to the observed data. For carrying out the refinement further, and for other temperatures, the magnetic moment along $c$-axis was allowed to refine. As the refinement progresses well only when considering the k-vector as incommensurate, the propagation vector, $\mathbf{k_y}$, was also refined for each temperature to find out its temperature dependence. It may be noted here that the the representation analysis is valid for (0, **k**, 0) with 0 < **k** < 1. Based on the symmetry elements used, the magnetic space group, identified using the Bilbao Crystallographic Server, is $C$2 (No. 5.13) in BNS setting [16-18].

The average magnetic moment on Tb per unit cell obtained by Rietveld fitting, as expected, is found to undergo a sudden increase at $T_N$ with lowering temperature with a slow variation and a tendency to saturate to about 5 $\mu_B$ below about 15 K, as shown in the inset of figure 1. Some projections of the magnetic structure at 4.5 K are shown in figures 3a (viewed along $c$-axis) and figure 3b (for three adjacent crystallographic unit cells along $b$-axis). It is found that the antiferromagnetic structure is of an incommensurate type with a propagation vector ($k$) of ± (0, $k_y$, 0), running along $b$-axis, as shown in figure 3c. If one moves along $b$-direction from one cell to the other, as shown in figure 3b, the Tb ions are aligned in the same direction for a given pair of Tb ions, but the direction of the pairs keep changing as 'up-up-down-down-down-down-up-up-down-down- down-down-up-up' with a change in the amplitude as shown schematically in figures 3b and 3c. It is noteworthy that $b$-component, $k_y$, after showing a sharp increase from 0.31 in the close vicinity of $T_N$, tends to stay near 0.33 below 17 K (Figure 4), as though there is a dramatic change in the magnetism with decreasing temperature. This is the same temperature at which glassy anomalies get triggered in the magnetization data [9]. A closer inspection of figure 4 however reveals that there is in fact a marginal decrease below about 12 K, which indicates additional magnetic complexities to be understood. In order to see any loss of intensity due to possible spin-glass behavior well below $T_N$ (as inferred from magnetization data), we have obtained integrated intensity of some magnetic Bragg peaks, which are plotted in figure 5. It is clear that a monotonic increase of these intensities as the temperature is lowered below $T_N$ is only observed, in contrast to expectations.

From the results reported in the literature [9], it is clear that the compound, $Tb_3Ru_4Al_{12}$, apparently undergoes long-range antiferromagnetic ordering at 22 K down to low temperatures, as inferred from the neutron diffraction data. It may be recalled [9] that the spin-glass anomalies were reported in dc χ, ac χ, and isothermal remnant magnetization data at temperatures lower than $T_N$. These measurements were performed on well-characterized single crystals and it is found that these spin-glass features appear for the orientation of the $c$-axis of the crystal along the magnetic field direction. This reveals that the glassy features are intrinsic to the compound (characterizing it as an anisotropic spin-glass) and not due to any magnetic impurity [19]. It is therefore puzzling that we are not able to see the loss of intensity of magnetic Bragg peaks in neutron diffraction pattern as one enters glassy phase and the incommensurate antiferromagnetic structure is apparently maintained even at 4.5 K. It is at present not clear to us how to reconcile these contradictory results from the bulk and neutron diffraction measurements. One possible explanation is that this system does not have a single propagation vector but one, which is varying, with respect to temperature below ~17 K, with the magnetic structure varying with time. We have also recorded the neutron diffraction patterns at several time intervals over a period of a few hours at many temperatures, but we are not able to resolve any time dependence of the diffraction pattern. However, it is to be noted that there is a qualitative change in the nature of the curve in figure 4 at 17 K and that there is a marginal dip around 10 K, as though the propagation vector behavior, as the temperature is lowered across 17 K, is a bit complex. Frequency dependent measurements, like ac $\chi$, would naturally reflect the dynamics of these propagation vectors. In other words, the observed glassiness in the bulk data can be attributed to the phenomenon of 'dynamic antiferromagnetic ordering' or 'spin-glass in time-domain' due to degeneracy of marginally different wave vectors, presumably facilitated by geometrical frustration. In other words, antiferromagnetism and spin-glass behavior are not conventional ones. At this juncture, it may be recalled that Chandra *et al* [20] proposed that the glassy features in geometrically frustrated systems is intrinsic, but not due to disorder. It was also theorized that the spin-glass dynamics could be unconventional and that kagome lattices should be the ideal testing grounds for this idea. There was also a recent proposal of 'dynamical spin-glass phenomenon' in magnetism [21]. Though there seems to be a consensus in the literature that disorder is required to trigger glassiness due to geometrical frustration, it thus appears that this question seems to be still open. We therefore wonder whether the compound under study is a model system for such proposals [20, 21]. Such a phenomenon is not unrealistic, as the phenomenon of slow order-order transition phenomenon has been demonstrated [22] for another geometrically frustrated system, $Ca_3Co_2O_6$ [23, 24]. This is in some sense another kind of spin-liquid, considering that the proposed fluctuation is a co-operative (antiferromagnetic) effect, though the terminology 'spin-liquids' in the past literature refers to non-magnetic ground state.

4. **Conclusion**

We have presented the results neutron diffraction measurements of a metallic kagome lattice, $Tb_3Ru_4Al_{12}$, which was previously identified to show reentrant spin-glass anomalies in the bulk data. The present results suggest that the magnetic structure is of a long-range antiferromagnetic type, however, with its propagation vector showing a complex temperature dependence. The results indicate complexity of the spin-dynamics of this compound. Therefore, further studies on this compound would be rewarding to advance the knowledge on manifestations of geometrically frustrated magnetism. It is also of interest to explore whether alternating small and big triangular Tb clusters, characteristic of this distorted kagome layer, are responsible for the observed magnetic behavior.

**Figures and Captions**

**Figure 1**

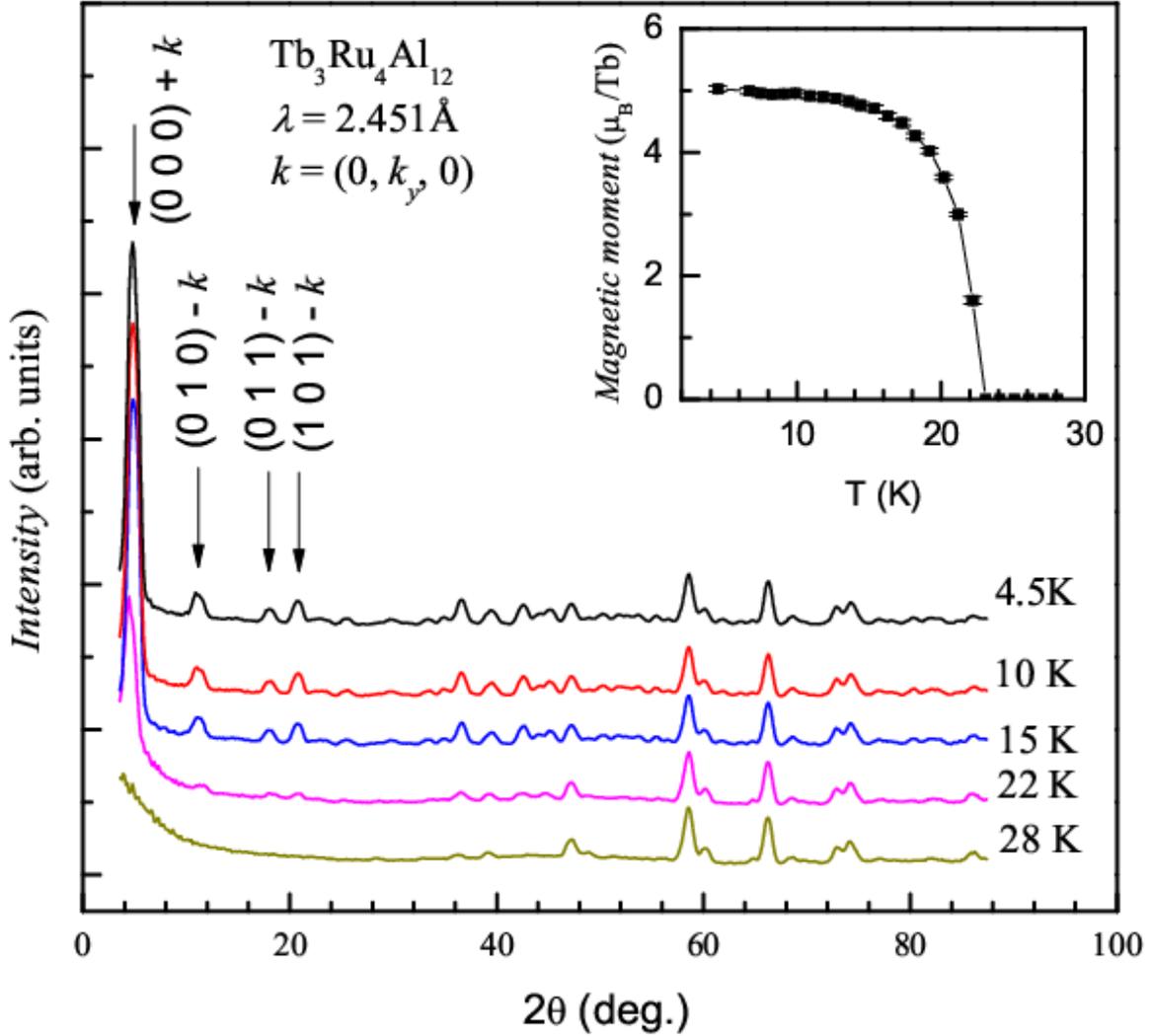

Figure 1: Neutron diffraction pattern of powdered $Tb_3Ru_4Al_{12}$ at selected temperatures. Vertical arrows mark magnetic Bragg peaks. The intensity is in arbitrary units, as the curves for each temperature are shifted with respect to each other for the sake of clarity. Inset shows how the average magnetic moment on Tb evolves with temperature below $T_N$.

**Figure 2**

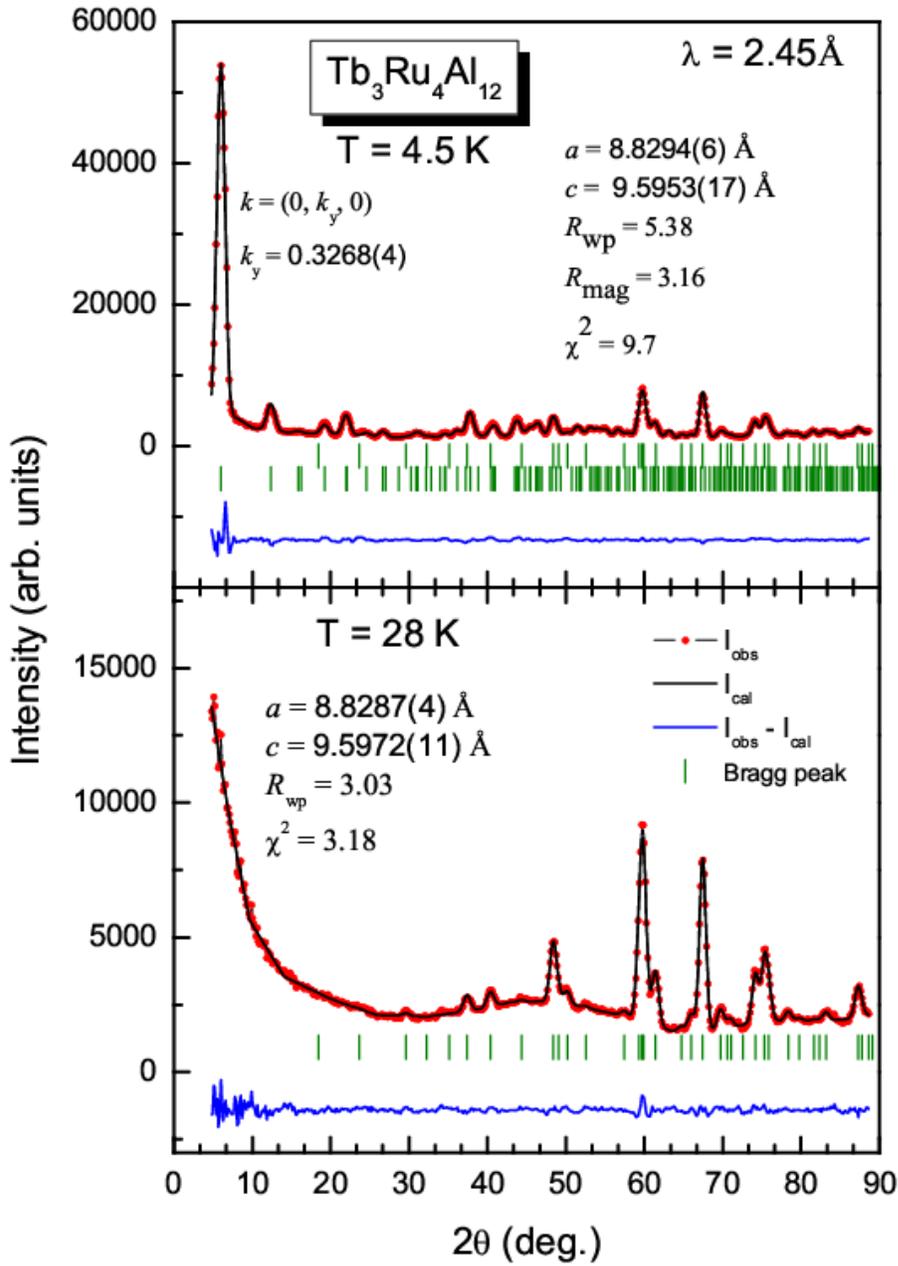

Figure 2: Rietveld refinement of the neutron diffraction pattern is shown for two temperatures, 4.5 K ($T < T_N$) and 28 K ($T > T_N$). The observed profile is shown as full red circles, calculated profile as continuous black line, and the difference curve as continuous blue line. The Bragg peak positions (nuclear on the top row and magnetic at the bottom row) are shown by vertical tick marks.

**Figure 3**
Figure 3(a)

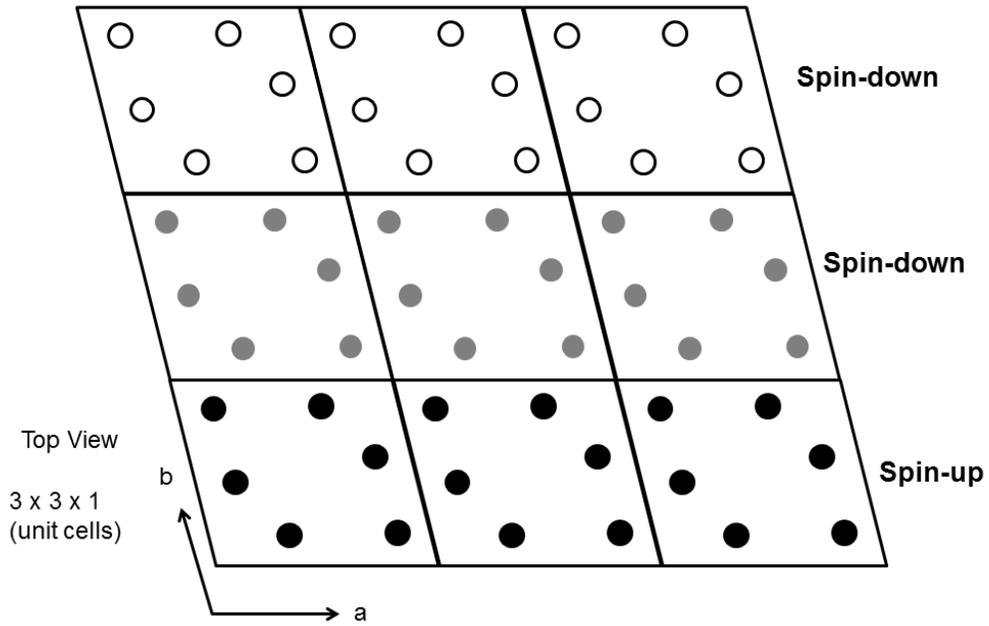

Figure 3(b)

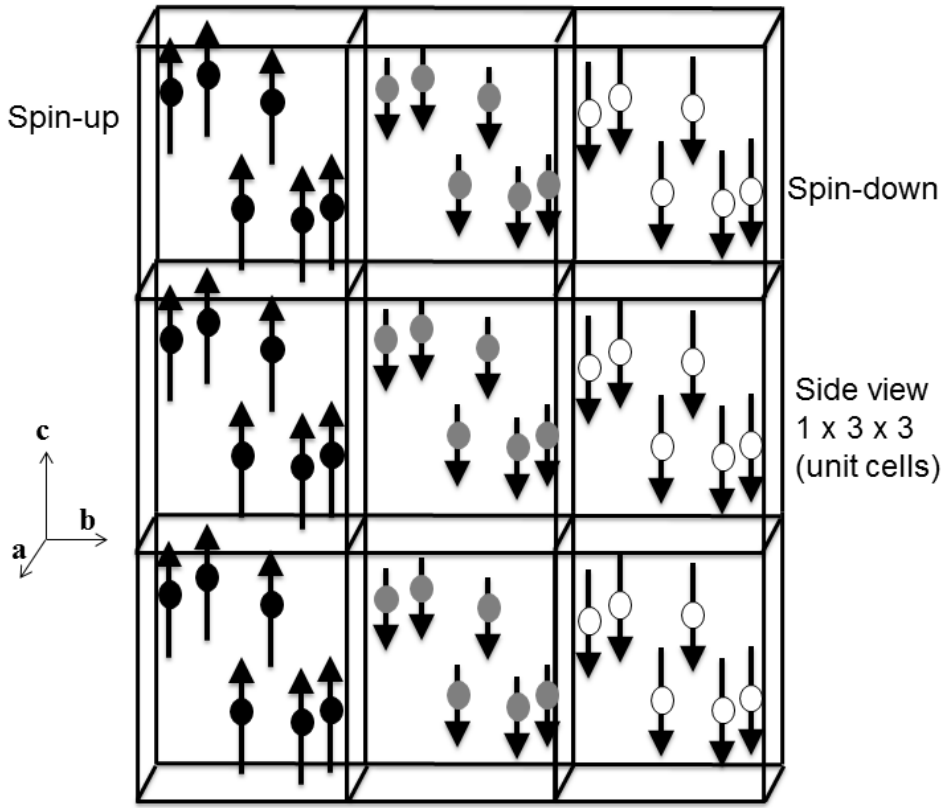

Figure 3(c)

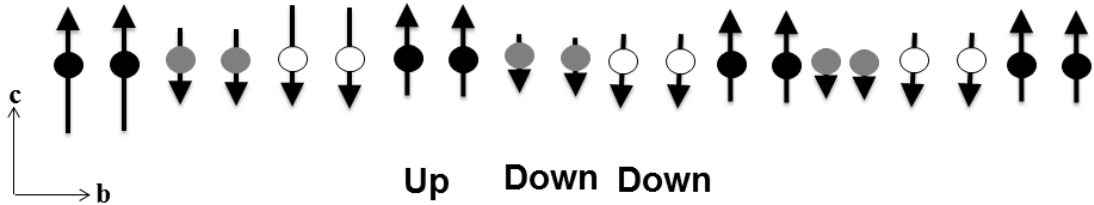

Figure 3: The schematic diagrams of (a) basal plan view (along $c$-direction) of the magnetic structure, (b) relative alignment of Tb ions in three adjacent unit cells along $b$-axis, and (c) modulation of the moments according to the propagation vector ($\mathbf{k}_y = 0.327$) along $b$-axis at 4.5 K for powdered $Tb_3Ru_4Al_{12}$ is schematically shown.

**Figure 4**

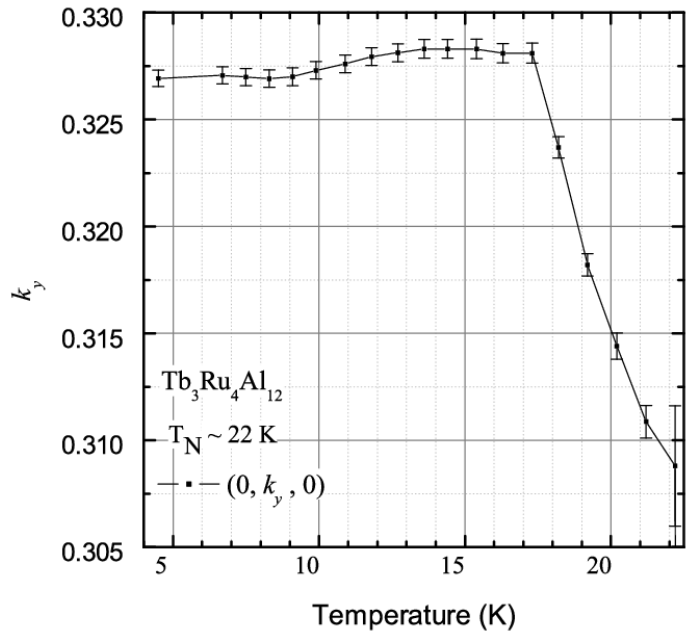

Figure 4: The temperature dependence of propagation vector along $b$-axis, $k_y$, is plotted.

**Figure 5**

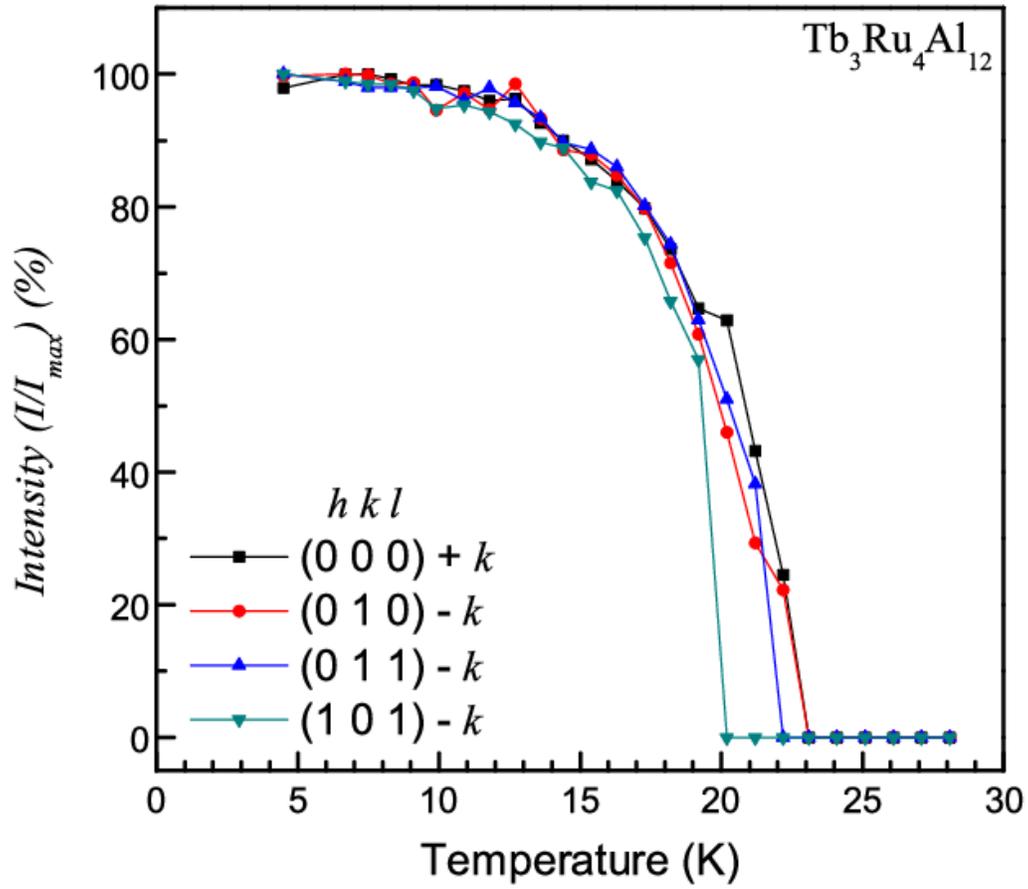

Figure 5: Temperature dependence of the integrated intensity of selected magnetic Bragg peaks of Tb$_3$Ru$_4$Al$_{12}$.

# Supplementary Information

# Neutron diffraction study of a metallic kagome lattice, $Tb_3Ru_4Al_{12}$

**Sudhindra Rayaprol,[1] Andreas Hoser,[2] Kartik K Iyer,[3] Sanjay K Upadhyay,[3] E. V. Sampathkumaran,[3]**

[1]UGC-DAE-Consortium for Scientific Research, Mumbai Centre, BARC Campus, Trombay, Mumbai – 400085, India
[2]Helmholtz-Zentrum Berlin für Materialien und Energie, Berlin, D-14109, Germany
[3]Tata Institute of Fundamental Research, Homi Bhabha Road, Colaba, Mumbai 400005, India

Here, we provide details of atomic positions (Table S1), reliability factors for Reitveld refinement of temperature dependent neutron diffraction profiles (Table S2), variation of structural parameters as a function of temperature (Table S3), supplementary information for magnetic structure refinement, propagation vector group information, data about magnetic cell, and basis vectors for propagation vector

**Table S1**: Atomic positions of the crystallographic unit cell of $Tb_3Ru_4Al_{12}$ at 28K

| Atom | Label | Wyckoff Site | Atomic Position | Occupancy |
|---|---|---|---|---|
| **Tb** | Tb | 6h | ($x$, 2$x$, 1/4) | 1.00 |
|  |  |  | $x = 0.19630(68)$ |  |
| **Ru** | Ru1 | 6g | (1/2, 0, 0) | 1.00 |
|  | Ru2 | 2a | (0, 0, 0) | 1.00 |
| **Al** | Al1 | 12j | ($x$, $y$, 1/4) | 0.54 |
|  |  |  | $x = 0.54892(553)$ |  |
|  |  |  | $y = 0.41435(508)$ |  |
|  | Al2 | 4f | (1/3, 2/3, $z$) | 1.00 |
|  |  |  | $z = 0.00635(169)$ |  |
|  | Al3 | 12k | ($x$, 2$x$, $z$) | 0.94 |
|  |  |  | $x = 0.15931(109)$ |  |
|  |  |  | $z = 0.58368(158)$ |  |
|  | Al4 | 2b | (0, 0, 1/4) | 0.97 |

**Table S2**: Reliability factors for Reitveld refinement of temperature dependent neutron diffraction profiles of $Tb_3Ru_4Al_{12}$

| Temperature (K) | $R_p$ | $R_{wp}$ | $R_{exp}$ | $\chi^2$ | Bragg-R | Rf-factor | $R_{Mag}$ |
|---|---|---|---|---|---|---|---|
| 4.50 | 4.52 | 5.38 | 1.73 | 9.70 | 4.83 | 2.93 | 3.16 |
| 6.70 | 4.85 | 5.59 | 1.73 | 10.4 | 4.82 | 2.78 | 4.96 |
| 7.50 | 5.28 | 5.55 | 1.73 | 10.2 | 4.79 | 2.83 | 5.82 |
| 8.30 | 5.32 | 5.67 | 1.73 | 10.7 | 4.94 | 3.09 | 5.42 |
| 9.10 | 4.93 | 5.60 | 1.73 | 10.4 | 4.91 | 2.96 | 5.25 |
| 9.90 | 5.26 | 5.52 | 1.73 | 10.1 | 4.61 | 2.78 | 6.28 |
| 10.9 | 5.31 | 5.62 | 1.74 | 10.5 | 4.74 | 2.86 | 5.96 |
| 11.8 | 5.44 | 5.62 | 1.73 | 10.5 | 4.70 | 2.80 | 5.96 |
| 12.7 | 5.48 | 5.60 | 1.73 | 10.4 | 4.64 | 2.79 | 6.91 |
| 13.6 | 5.45 | 5.64 | 1.73 | 10.6 | 4.81 | 2.91 | 6.25 |
| 14.4 | 5.25 | 5.63 | 1.73 | 10.6 | 4.62 | 2.94 | 6.24 |
| 15.4 | 5.50 | 5.47 | 1.73 | 10.0 | 4.59 | 2.87 | 6.64 |
| 16.3 | 5.12 | 5.33 | 1.73 | 9.46 | 4.54 | 2.82 | 6.21 |
| 17.3 | 5.52 | 5.21 | 1.73 | 9.03 | 4.41 | 2.72 | 7.61 |
| 18.2 | 6.08 | 4.79 | 1.73 | 7.64 | 3.86 | 2.61 | 9.28 |
| 19.2 | 5.35 | 4.35 | 1.74 | 6.27 | 3.56 | 2.53 | 11.3 |
| 20.2 | 5.73 | 4.10 | 1.73 | 5.58 | 3.62 | 2.53 | 15.6 |
| 21.2 | 4.34 | 3.52 | 1.74 | 4.10 | 3.44 | 2.62 | 16.2 |
| 22.2 | 6.06 | 3.77 | 1.74 | 4.70 | 2.88 | 2.42 | 52.1 |
| 23.1 | 2.48 | 2.96 | 1.73 | 2.93 | 4.41 | 3.60 | -- |
| 24.1 | 2.20 | 2.83 | 1.73 | 2.67 | 4.60 | 3.92 | -- |
| 25.1 | 2.45 | 2.75 | 1.66 | 2.74 | 5.83 | 5.02 | -- |
| 26.1 | 2.52 | 2.83 | 1.66 | 2.90 | 5.66 | 5.00 | -- |
| 27.1 | 2.57 | 2.82 | 1.66 | 2.88 | 5.56 | 4.24 | -- |
| 28.1 | 2.41 | 3.03 | 1.70 | 3.18 | 4.45 | 5.40 | -- |

**Table S3**: Variation of structural parameters as a function of temperature for $Tb_3Ru_4Al_{12}$

| T (K) | $a = b$ (Å) | $c$ (Å) | Volume (Å$^3$) | $\mu_{Tb}$ ($\mu_B$) | $k_y$ |
|---|---|---|---|---|---|
| 4.50 | 8.82944(62) | 9.59537(172) | 647.827(133) | 5.030(49) | 0.32689(39) |
| 6.70 | 8.82920(65) | 9.59535(178) | 647.790(138) | 4.996(50) | 0.32706(40) |
| 7.50 | 8.82869(65) | 9.59581(178) | 647.746(137) | 4.959(50) | 0.32699(40) |
| 8.30 | 8.82892(66) | 9.59587(182) | 647.784(141) | 4.940(50) | 0.32690(41) |
| 9.10 | 8.82922(65) | 9.59562(180) | 647.811(139) | 4.946(50) | 0.32701(41) |
| 9.90 | 8.82878(65) | 9.59608(178) | 647.777(138) | 4.952(50) | 0.32731(40) |
| 10.9 | 8.82898(66) | 9.59559(181) | 647.774(140) | 4.915(51) | 0.32765(41) |
| 11.8 | 8.82919(66) | 9.59530(181) | 647.785(140) | 4.899(51) | 0.3279(42) |
| 12.7 | 8.82936(66) | 9.59524(180) | 647.806(140) | 4.878(51) | 0.32814(42) |
| 13.6 | 8.82856(67) | 9.59675(183) | 647.791(142) | 4.824(51) | 0.32838(43) |
| 14.4 | 8.82934(68) | 9.59628(183) | 647.874(142) | 4.762(52) | 0.32835(45) |
| 15.4 | 8.82892(66) | 9.597219(179) | 647.872(139) | 4.711(51) | 0.32836(45) |
| 16.3 | 8.82935(65) | 9.59497(174) | 647.786(135) | 4.591(49) | 0.32818(45) |
| 17.3 | 8.82885(64) | 9.59569(172) | 647.762(134) | 4.481(49) | 0.32817(47) |
| 18.2 | 8.82932(60) | 9.59534(160) | 647.807(125) | 4.266(46) | 0.32369(49) |
| 19.2 | 8.82931(55) | 9.59591(146) | 647.844(114) | 4.024(43) | 0.31849(52) |
| 20.2 | 8.82924(54) | 9.59589(139) | 647.833(109) | 3.596(41) | 0.31443(61) |
| 21.2 | 8.82926(45) | 9.59683(114) | 647.898(90) | 2.994(35) | 0.31088(76) |
| 22.2 | 8.82944(50) | 9.59748(124) | 647.969(99) | 1.603(59) | 0.30881(270) |
| 23.1 | 8.82911(42) | 9.59826(104) | 647.937(83) | 0 | 0 |
| 24.1 | 8.82876(41) | 9.59798(98) | 647.902(79) | 0 | 0 |
| 25.1 | 8.82924(44) | 9.59809(106) | 647.981(85) | 0 | 0 |
| 26.1 | 8.82907(43) | 9.59794(105) | 647.946(84) | 0 | 0 |
| 27.1 | 8.82924(42) | 9.59662(102) | 647.882(82) | 0 | 0 |
| 28.1 | 8.82874(46) | 9.59721(115) | 647.854(91) | 0 | 0 |

**Supplementary Information for Magnetic Structure refinement obtained from the output of the BasIrep (BSR) file**

1. Space group: $P\,6_3/m\,m\,c$
2. The conventional k-vector is: (0.00000  0.33333  0.00000)
3. The Generators of the little group of Brillouin zone point: 00
4. The little group can be generated from the following 2 elements:-
   => GENk(1): -x+y,y,-z-1/2

 => GENk(2): x,y,-z-1/2

5. Representative elements of the little group of Brillouin zone point: 00

   Operator of Gk Number( 1): x,y,z

   Operator of Gk Number( 2): -x+y,y,-z-1/2

   Operator of Gk Number( 3): x,y,-z-1/2

   Operator of Gk Number( 4): -x+y,y,z

=> Number of elements of G_k:    4

=> Number of irreducible representations of G_k:    4

=> Dimensions of Ir(reps):   1  1  1  1

| Ireps | Symmetry Operators | | | |
|---|---|---|---|---|
| v | 1 | 2 x,2x,1/4 | m x,y,1/4 | m x,2x,z |
| v | {1 \| 000} | {2_x2x0 \| 00p} | {m_xy0 \| 00p} | {m_x2xz \| 000} |
| v | Symm (1) | Symm (2) | Symm (3) | Symm (4) |
| Γ(1) | 1 | 1 | 1 | 1 |
| Γ (2) | 1 | 1 | -1 | -1 |
| Γ (3) | 1 | -1 | 1 | -1 |
| Γ (4) | 1 | -1 | -1 | 1 |

## *Propagation Vector Group Information*

1. The input propagation vector k IS NOT equivalent to -k, therefore the extended little group is G(k,-k)

2. The operators following the k-vectors constitute the co-set decomposition G[Gk]
3. The list of equivalent k-vectors are also given on the right of operators.
=> Numerals of the extended little group operators: {1, 12, 15, 22, 3, 10, 13, 24 }
=> The star of k is formed by the following   6 vectors:

    k_1 = (0.0000  0.3333  0.0000 )   Op: ( 1) x,y,z
                           Op: ( 12) -x+y,y,-z+1/2   -> (  0.0000  0.3333  0.0000 )
                             Op: ( 15) x,y,-z+1/2       -> (  0.0000  0.3333  0.0000 )
                             Op: ( 22) -x+y,y,z         -> (  0.0000  0.3333  0.0000 )

    k_2 = (0.3333 -0.3333  0.0000 )   Op: ( 2) -y,x-y,z
                             Op: ( 11) x,x-y,-z+1/2    -> (  0.3333 -0.3333  0.0000 )
                             Op: ( 18) -y,x-y,-z+1/2   -> (  0.3333 -0.3333  0.0000 )
                             Op: ( 19) x,x-y,z         -> (  0.3333 -0.3333  0.0000 )

Eqv. -k: k_3 = (0.0000 -0.3333  0.0000 )   Op: ( 3) -x,-y,z+1/2
                             Op: ( 10) x-y,-y,-z       -> (  0.0000 -0.3333  0.0000 )
                             Op: ( 13) -x,-y,-z        -> (  0.0000 -0.3333  0.0000 )
                             Op: ( 24) x-y,-y,z+1/2    -> (  0.0000 -0.3333  0.0000 )

    k_4 = (0.3333  0.0000  0.0000 )   Op: ( 4) y,x,-z
                             Op: (  9) x-y,x,z+1/2     -> (  0.3333  0.0000  0.0000 )
                             Op: ( 17) x-y,x,-z        -> (  0.3333  0.0000  0.0000 )
                             Op: ( 20) y,x,z+1/2       -> (  0.3333  0.0000  0.0000 )

    k_5 = (-0.3333  0.0000  0.0000 )   Op: ( 5) -x+y,-x,z
                             Op: (  8) -y,-x,-z+1/2    -> ( -0.3333  0.0000  0.0000 )
                             Op: ( 16) -y,-x,z         -> ( -0.3333  0.0000  0.0000 )
                             Op: ( 21) -x+y,-x,-z+1/2  -> ( -0.3333  0.0000  0.0000 )

    k_6 = (-0.3333  0.3333  0.0000 )   Op: ( 6) y,-x+y,z+1/2
                             Op: (  7) -x,-x+y,-z      -> ( -0.3333  0.3333  0.0000 )
                             Op: ( 14) y,-x+y,-z       -> ( -0.3333  0.3333  0.0000 )
                             Op: ( 23) -x,-x+y,z+1/2   -> ( -0.3333  0.3333  0.0000 )

=> G_k has the following symmetry operators:
   1 SYMM( 1) = x,y,z
   2 SYMM( 12) = -x+y,y,-z+1/2
   3 SYMM( 15) = x,y,-z+1/2
   4 SYMM( 22) = -x+y,y,z

## *Data about magnetic cell*

The atom Tb at Wyckoff site *6h* is split in 4 orbits

Therefore, while calculating for axial vectors, the atoms within a primitive unit cell are:

| Atom | x | y | z | Symmetry |
|---|---|---|---|---|
| For Site : 1 | | | | |
| Tb1_1 | 0.1963 | 0.3926 | 0.2500 | (x, y, z) |
| For Site: 2 | | | | |
| Tb2_1 | 0.6074 | 0.8037 | 0.2500 | (x, y, z) |
| Tb2_2 | 0.1963 | 0.8037 | 0.2500 | (-x+y, y, -z+1/2) + (0, 0, 0) |
| For Site: 3 | | | | |
| Tb3_1 | 0.8037 | 0.6074 | 0.7500 | (x, y, z) |
| For Site: 4 | | | | |
| Tb4_1 | 0.3926 | 0.1963 | 0.7500 | (x, y, z) |
| Tb4_2 | 0.8037 | 0.1963 | 0.7500 | (-x+y, y, -z+1/2) + (1, 0, 1) |

The decomposition of the magnetic representations for Tb (6h) site can be written as

| Site | Gamma ($\Gamma_{magnetic}$) |
|---|---|
| 1 | 1 $\Gamma(2)$ + 1 $\Gamma(3)$ + 1 $\Gamma(4)$ |
| 2 | 1 $\Gamma(1)$ + 2 $\Gamma(2)$ + 1 $\Gamma(3)$ + 2 $\Gamma(4)$ |
| 3 | 1 $\Gamma(2)$ + 1 $\Gamma(3)$ + 1 $\Gamma(4)$ |
| 4 | 1 $\Gamma(1)$ + 2 $\Gamma(2)$ + 1 $\Gamma(3)$ + 2 $\Gamma(4)$ |

**Table S4**: The basis vectors for the propagation vector $k$ = (0, 0.3333, 0) for Wyckoff position *6h* for representation

| For Sites 1 and 3 | $\Gamma 2$ | $\Gamma 3$ | $\Gamma 4$ | | | |
|---|---|---|---|---|---|---|
| | BV1 | BV1 | BV1 | | | |
| x, y, z | 0.5 1 0 | 0 0 1 | 1 0 0 | | | |
| | 0 0 0 | 0 0 0 | 0 0 0 | | | |
| For Sites 2 and 4 | $\Gamma 1$ | $\Gamma 2$ | | $\Gamma 3$ | $\Gamma 4$ | |
| | BV1 | BV1 | BV2 | BV1 | BV1 | BV2 |
| x, y, z | 0 0 1 | 1 0 0 | 0 1 0 | 0 0 1 | 1 0 0 | 0 1 0 |
| | 0 0 0 | 0 0 0 | 0 0 0 | 0 0 0 | 0 0 0 | 0 0 0 |
| -x+y,y,-z+1/2 | 0 0 -1 | -1 0 0 | 1 1 0 | 0 0 1 | 1 0 0 | -1 -1 0 |
| | 0 0 0 | 0 0 0 | 0 0 0 | 0 0 0 | 0 0 0 | 0 0 0 |

- $\Gamma 3$ representation was used for all the sites for refining the magnetic phase of Tb$_3$Ru$_4$Al$_{12}$ sample.
- The Tb moments were constrained to be same at all the four sites to obtain the best fit.

**Figure S1**

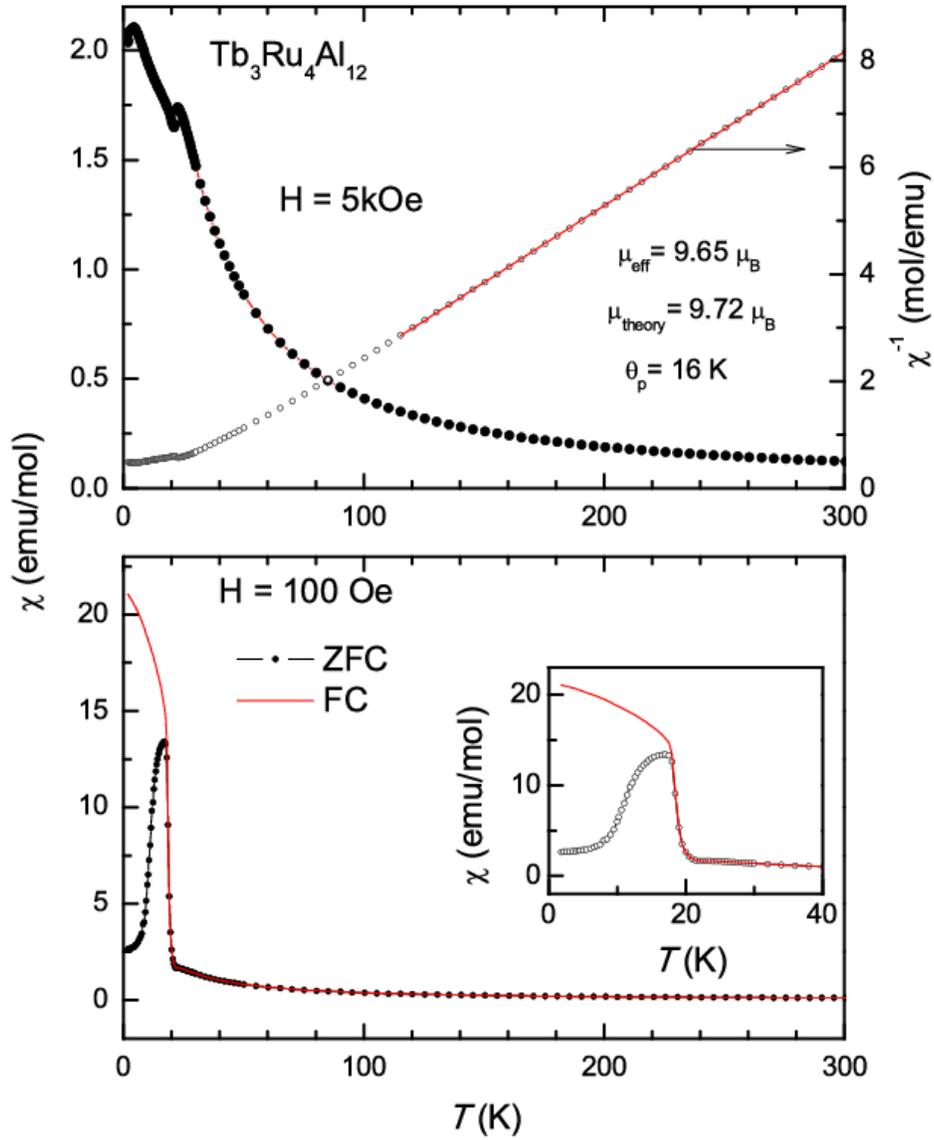

**Figure S1:** The magnetic susceptibility ($\chi = M/H$) for the polycrystalline sample of $Tb_3Ru_4Al_{12}$ is measured as a function of temperature in applied fields of 5 kOe (upper panel) and 100 Oe (lower panel). The features correspond quite well with those reported by us previously (see Ref. 9).